\pdfoutput=1
\documentclass[preprint,12pt]{elsarticle}

\usepackage{epsfig}
\usepackage{url}

\usepackage{amssymb}

\begin{document}

\begin{frontmatter}



\title{First operation of a double phase LAr Large Electron Multiplier Time Projection Chamber
with a two-dimensional projective readout anode}


\author{A.~Badertscher}
\author{A.~Curioni}
\author{L.~Knecht}
\author{D.~Lussi}
\author{A.~Marchionni}
\author{G.~Natterer}
\author{F.~Resnati}
\author{A.~Rubbia}
\author{T.~Viant}

\address{Institute for Particle Physics, ETH Zurich, 8093 Zurich, Switzerland}

\begin{abstract}

We have previously reported on the construction and successful operation of the novel double phase Liquid Argon Large Electron Multiplier Time Projection Chamber (LAr LEM-TPC). This detector concept provides a 3D-tracking and calorimetric device capable of adjustable charge 
amplification, a promising readout technology for next-generation neutrino detectors and direct Dark Matter searches.

In this paper, we report on the first operation of a LAr LEM-TPC prototype -- with an active area of 10$\times$10~cm$^2$ and 21~cm drift length --
equipped with a single 1~mm thick LEM amplifying stage and a two dimensional projective readout anode.
Cosmic muon events were collected, fully reconstructed and used to characterize the performance of the chamber.
The obtained signals provide images of very high quality and
the energy loss distributions of minimum ionizing tracks give a direct estimate of the amplification. 
We find that a stable gain of 27
can be achieved with this detector configuration corresponding to a signal-over-noise ratio larger than 200 for minimum ionizing 
tracks. The decoupling of the amplification stage and the use of the 2D readout anode offer several advantages which are
described in the text.
\end{abstract}


\end{frontmatter}


\section{Introduction}

The Liquid Argon Large Electron Multiplier Time Projection Chamber (LAr LEM-TPC) is 
a novel kind of double phase (liquid-vapor) noble gas TPC with adjustable gain~\cite{Badertscher:2008rf}. 
The chamber is a precise tracking device that allows to
reconstruct three-dimensionally the position and the morphology of
ionizing tracks,  $dE/dx$ information with high sampling rate, and it acts as high resolution calorimeter for contained events.
Thanks to the gain, each volumetric pixel (voxel) is reconstructed 
with high signal-to-noise ratio and consequently a low energy deposition threshold is possible.
In addition charge amplification 
reduces the impact of charge dilution due to the longitudinal diffusion of the electron cloud along the drift paths, 
and can be used to compensate for potential charge losses due to electronegative impurities diluted in the liquid argon. 
We believe that this technology is very promising
for the realization of next generation underground
detectors for neutrino physics and proton decay searches~\cite{Rubbia:2004tz,Rubbia:2009md,Badertscher:2010sy}
and for direct Dark Matter detection~\cite{Rubbia:2005ge} with imaging technique, in all cases with
a significant improvement of the imaging quality compared to 
the single phase liquid argon TPC~\cite{Badertscher:2009av}.

In this paper, we report on the first operation of a LAr LEM-TPC prototype with an active area of 10$\times$10~cm$^2$ and 21~cm drift length,
equipped with a single 1~mm thick LEM amplifying stage and a two dimensional projective readout anode.

The charge produced by ionizing particles in the liquid is drifted towards the liquid-vapor interface, where electrons are extracted 
to the vapor phase by means of an appropriate
electric field produced by two grids. In the vapor phase, Townsend avalanche takes place in the high electric field regions confined 
in the LEM holes, similar to the situation of the Gas Electron Multiplier (GEM)~\cite{Sauli:1997qp}. 
The LEM is a macroscopic hole electron multiplier built with standard PCB techniques. 
The amplified charge is collected by a set
of segmented electrodes, on which signals are induced.

In order to obtain a complete 3D-spatial reconstruction of ionizing events, moving charges have to induce signals on (at least) two complementary X-Y
sets of electrode strips. The Z~(drift) coordinate
is extracted from the drift time determined from the $T_0$ given by immersed DUV-sensitive photo detectors recording the prompt
LAr scintillation light.
In our previous setups~\cite{Badertscher:2009av,Badertscher:2008rf,Badertscher:2010fi},  we implemented two perpendicular 
X\&Y views by a segmentation of the upper electrode of the LEM (X-view) and of the anode (Y-view) with 6~mm pitch. 
However, long-term operation in stable conditions with gain indicated
that a fine segmentation of the LEM creates electric field distortions which may trigger discharges. We
concluded on the necessity to decouple the charge multiplication and readout stages~\cite{Badertscher:2010fi}.

This separation has been achieved thanks to the use of a projective 2D charge X-Y-readout, providing two independent views with 3~mm effective pitch. 
Based on the GEM 2D readout concept~\cite{Bressan:1998jj}, it consists of two perpendicular sets of strips with a 50~$\mu$m thick Kapton spacer in-between. The drifting charge is collected on both sets of strips, based on the principle of charge sharing. 

This new layout has several advantages: (1) the functions of amplification and of readout are decoupled, i.e.
avoiding the need for a segmented LEM plane, which is more prone to electric discharges; (2) as a consequence, there is no need
to install HV decoupling capacitors on the LEM electrodes; (3) the pitch of the readout is not constrained by the LEM hole size and pitch;
(4) the amplified charge is effectively collected on a single plane where two perpendicular sets of strips of a two dimensional projective 
anode are located, simplifying the overall design; (5) with a proper strip geometry, the charge can be equally shared and collected on both readout views,
therefore the signal shapes are unipolar and identical on both views, easing the reconstruction and feature extraction of the signal
waveforms.

In section~\ref{sec:detector} we briefly describe the experimental setup. The chosen electric configuration 
is motivated in section~\ref{sec:driftfield}.
LEM and projective anode are then detailed in the sections~\ref{sec:LEM} and \ref{sec:anode}. Section~\ref{sec:operation}
describes the operation of the chamber, while cosmic muon reconstruction is presented in section ~\ref{sec:reco}.
Results on the characterization of the detector are given in section~\ref{sec:results}.

\begin{figure}[tbp]
   	\centering
   	\includegraphics[width=0.75\textwidth]{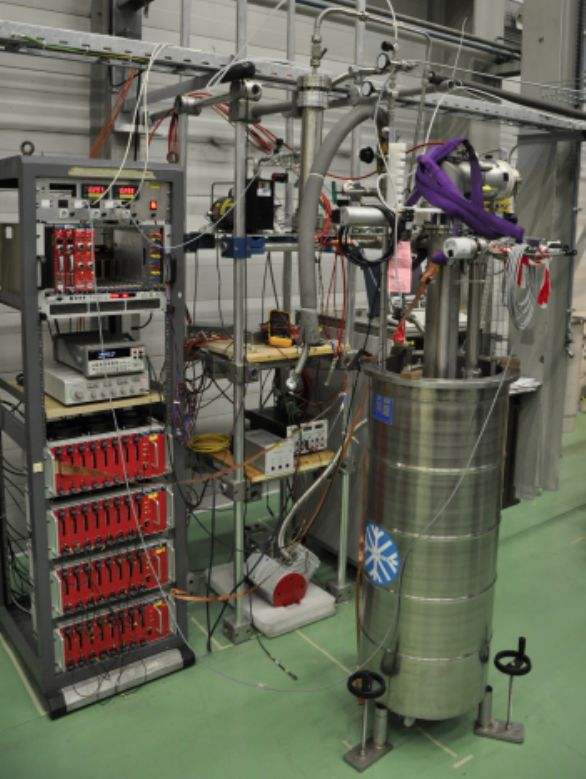} 
   	\caption{Overview of the 3L setup.}
   	\label{fig:overalllayout}
\end{figure}

\section{The experimental setup}
\label{sec:detector}
\begin{figure}[htbp]
   	\centering
   	\includegraphics[height=10cm]{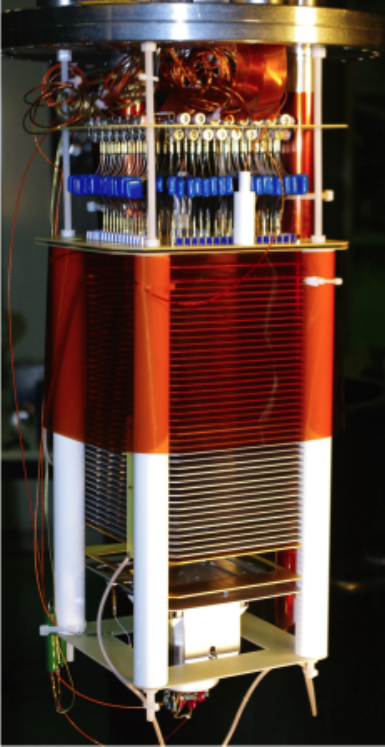} 
	\hspace*{2cm}
    	\includegraphics[height=10cm]{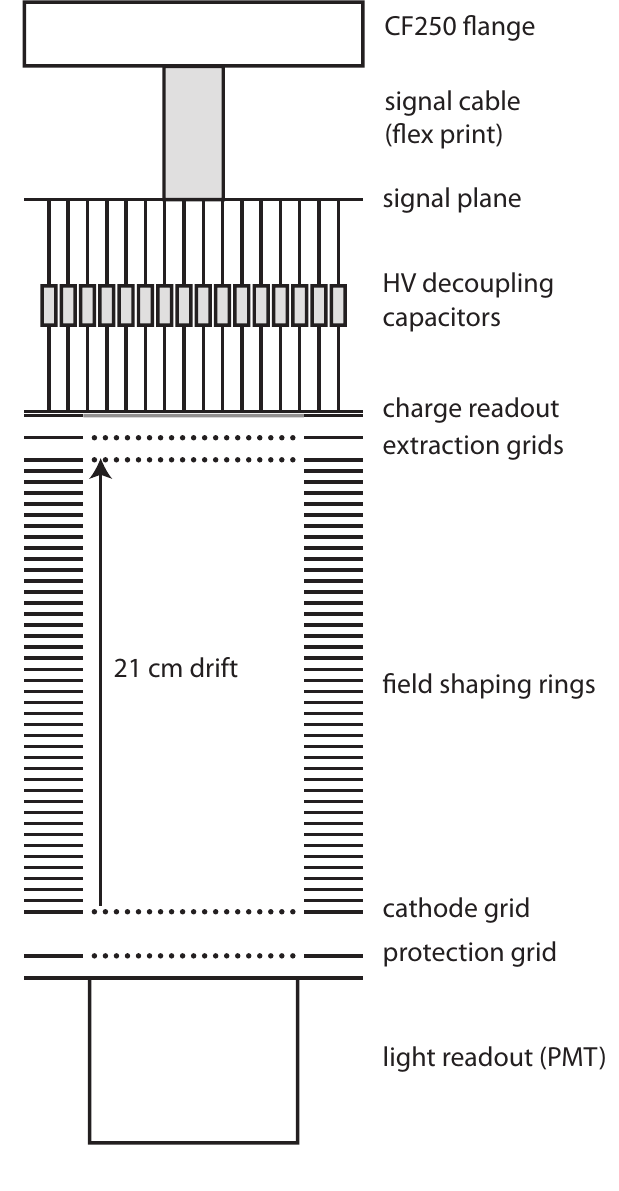} 
   	\caption{Left: picture of the 3L prototype LEM-TPC. Right: schematics of a LEM-TPC.}
   	\label{fig:detector}
\end{figure}

\begin{figure}[htbp]
   	\centering
    	\includegraphics[width=0.75\textwidth]{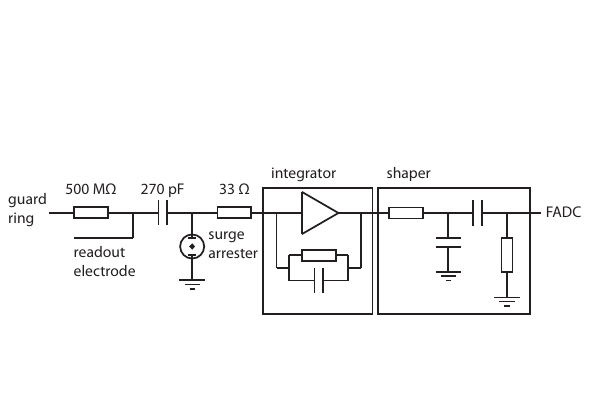} 
   	\caption{Electrical schematics of each readout channel of the anode.}
   	\label{fig:elecscheme}
\end{figure}

The purpose of this section is to briefly describe the 3L prototype chamber used as a test-bench setup (see Figure~\ref{fig:overalllayout}). A more detailed description can be found elsewhere~\cite{Badertscher:2008rf}. Figure~\ref{fig:detector} shows a picture of the detector (left), with the corresponding schematics (right). The LAr drift volume has a cross-section of 10x10 cm$^2$ and the distance between cathode and LAr surface equals 21~cm. Below the cathode grid, optically transparent  with a wire pitch of 5 mm, a Hamamatsu R6237-01 photomultiplier tube, coated with the wavelength shifter tetraphenylbutadiene (TPB)~\cite{Boccone:2009kk} is positioned below a grounded protection grid. It allows to detect the liquid argon scintillation light, peaked around 128 nm~\cite{Takahashi1983591}. The uniform drift field is defined by equally-spaced rectangular field shaping rings, connected to a chain of 940~M$\Omega$ resistors. On top of the drift volume there are two parallel extraction grids with a wire pitch of 5 mm and a gap of 10 mm. These two grids allow to apply an electric field across the LAr surface which is
independent of the chosen drift field. The leveling of the LAr is done with millimeter precision by a capacity measurement of the two grids. The LEM and the 2D anode are mounted 1~cm above the upper grid and 2 mm apart from each other. It is worth mentioning that the 2D anode can be operated at high voltage. This is done 
(see Figure~\ref{fig:elecscheme}) by connecting each strip with a 500~M$\Omega$ resistor to the guard ring.  
HV capacitors (270~pF) decouple the fast signals from the high voltage. Surge arresters\footnote{EC 90, EPCOS AG, Munich, Germany} are used to protect the readout electronics from discharges. 
In order to measure the signals, custom-made low-noise JFET charge preamplifiers with a integration time constant of 470 $\mu$s
were developed with a design derived from Ref.~\cite{1344263}.
The charge integrator has four low noise 
BF862~FET transistors from Philips Semiconductor\footnote{Now available at \url{http://www.nxp.com}.} connected in parallel to match a high detector capacitance. 
Its charge sensitivity is 1~mV/fC, as determined by the 1~pF feedback capacitance. The integrator stage is followed 
by a RC-CR shaper with a gain of about 10 and shaper integration and differentiation time constants of 0.6~$\mu$s and 2~$\mu$s, respectively. 
The amplifiers were designed to be compatible with both positive and negative 
inputs and are provided with an input for the adjustment of the output baseline, in order to utilize the full dynamic range 
of the digitizer.
Two preamplifier channels are housed on a single hybrid. 
We measured a sensitivity of about 12~mV/fC and a signal to noise ratio of 10 for 1~fC input charge and 200~pF input capacitance.
The digitization and data acquisition is done by the CAEN SY2791 readout system\footnote{CAEN S.p.A., Viareggio, Italy, \url{http://www.caen.it}}, 
specially developed as 
a complete readout system for LAr TPCs, housing up to 8  boards with 32 channels each, 
with one preamplifier per channel, 2.5 MHz 12-bit flash ADCs, trigger logic, and an optical link.

\section{Electric fields configuration}
\label{sec:driftfield}
\begin{figure}[htbp]
   	\centering
   	\includegraphics [width=0.7\textwidth]{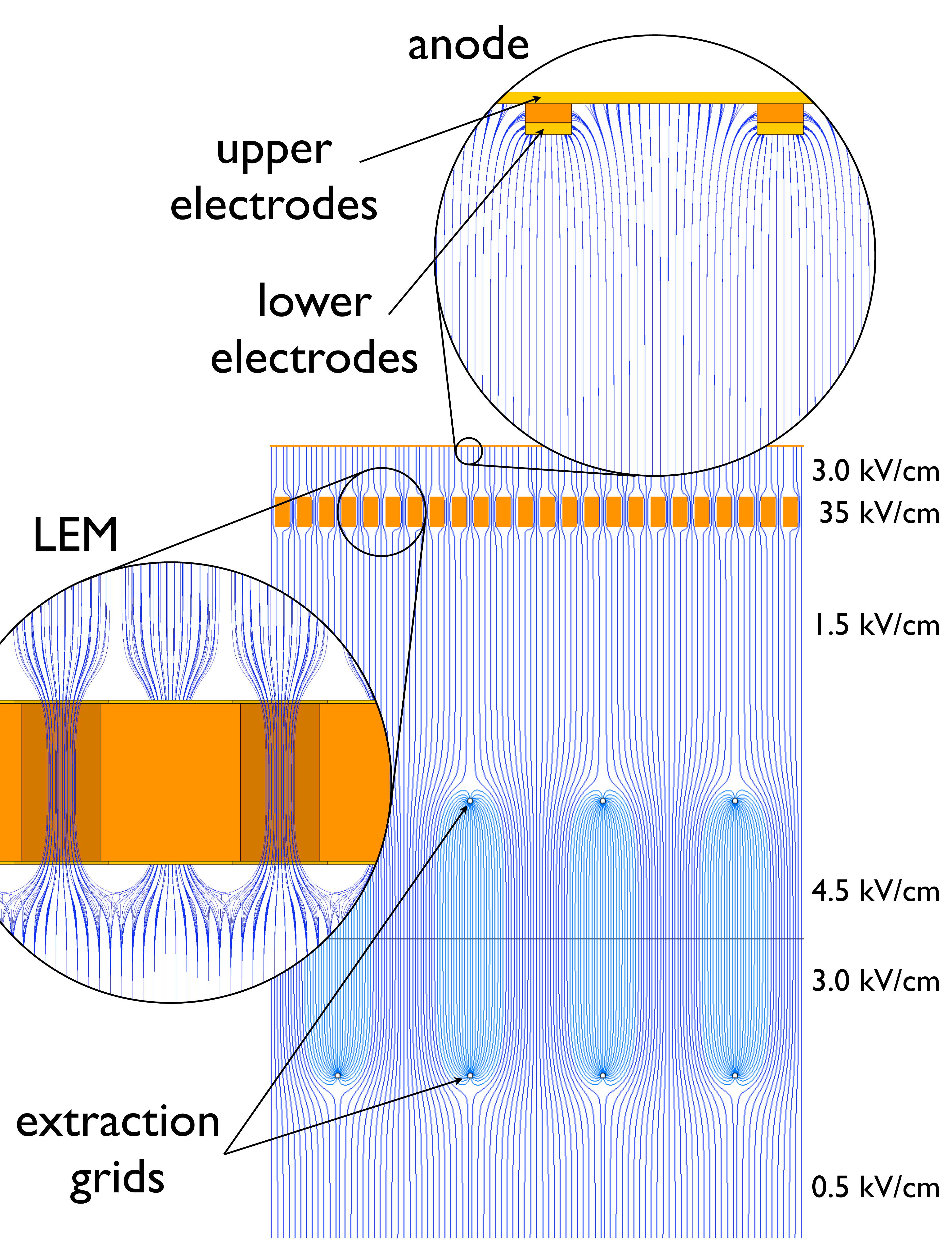}
   	\caption{Overall electric field configuration including (a) drift region (b) extraction from liquid to vapor (c) drift towards LEM holes (d)
	LEM holes (e) drift towards anode (f) fields near the anode strips. The displayed configuration corresponds to equal charge sharing between
	X- and Y-views.}
	\label{fig:inductionField}
\end{figure}

The electric field configuration is constrained by a few requirements: (a)
the drift field is set to 500~V/cm, which corresponds to a drift velocity of 1.63~mm/$\mu$s~\cite{Walkowiak:2000wf}
and the electron-ion-pair recombination factor is 70\%~\cite{Amoruso:2004dy} for minimum ionizing tracks;
(b)  in order to extract electrons fast and efficiently from the liquid to the vapor phase, 
extraction fields larger than 2.5 kV/cm are applied~\cite{Gushchin1982,Borghesani1990481}; 
(c) the field inside the 1~mm thick LEM holes must reach 35.5~kV/cm (see section~\ref{sec:LEM});
(d) the strengths of the transfer fields must minimize the use of high voltages.
With such fields, the charge produced in the liquid must be drifted, extracted and efficiently collected on the anode plane.
The electric charges will traverse several regions with
different electric fields. The transparency of the system drift-extraction grid-LEM-anode must be maximized by appropriate
choice of the transfer fields. In addition, the amount of charge
collected on the X- and Y-views of the anode depends on their potentials and geometry. 
Therefore, the electric field configuration of the entire system has been numerically simulated and optimized, using 
COMSOL Multiphysics\footnote{COMSOL multiphysics, \url{http://www.comsol.com}}. 
The typical configurations of the electric fields are reported in Table~\ref{fieldConfiguration}.
The charge is emitted into the gas phase by means of a 3 kV/cm extraction field. 
In order to maximize the LEM transparency, the field between the extraction grid in GAr and 
the bottom LEM electrode is reduced down to 1.5~kV/cm and the field between 
LEM and Anode is equal to 3~kV/cm. 
The corresponding voltage applied across the two electrodes of the 1~mm thick LEM is as high as 3.55 kV. 
The two anode views are kept at the same potential and their geometries are defined to collect
equal amounts of charge.
Figure~\ref{fig:inductionField} shows the computed electric field lines. The effect of the field line squeezing
in the holes is shown in the corresponding zoomed area. The impact of the extraction grids on the field lines
is also visible. We note that because of the changing dielectric constant, the field above the liquid is
$1.5\times$ stronger than in the liquid.  

\begin{table}[h]
\begin{minipage}[b]{18pc}
\begin{center}
\begin{tabular}{lr}
  \hline
  \hline
  Anode-LEM & 3~kV/cm\\
  LEM holes & 30-35.5~kV/cm\\
  Grid-LEM & 1.5~kV/cm\\
  Extraction (in LAr) & 3~kV/cm\\
  Drift & 500~V/cm\\
  \hline
  \hline
\end{tabular}
\end{center}
\end{minipage}
\hfill
\begin{minipage}[b]{14pc}
\caption{\label{fieldConfiguration}Nominal configuration of electric fields.}
\end{minipage}
\end{table}

\section{The Large Electron Multiplier (LEM)}
\label{sec:LEM}
The general technique of electron multiplication via avalanches in small holes
is attractive because (1)~the required high electric field can be naturally attained
inside the holes and (2)~the finite size of the holes effectively ensures 
a confinement of the electron-ion pair avalanche, thereby reducing secondary 
effects in a medium without quencher. 
The gain ($G$) in a given uniform electric field of a parallel plate chamber  
at a given pressure is described by $G\equiv~e^{\alpha d}$ where 
$d$ is the gap thickness and  $\alpha$
 is the first Townsend coefficient, which represents the number of electrons
 created per unit path length by an electron in the amplification region. 
The behavior of this coefficient 
with density and electric field can be approximated by the Rose and 
Korff law~\cite{RoseKorff}: $\alpha(E)=A\rho~e^{-B\rho~/E}$
where  $E$ is the electric field, 
$\rho$ is the gas density, $A$ and $B$ are the parameters depending on the gas.
A fit to the electric field dependence of the Townsend
coefficient in the range between 20 and 40~kV/cm predicted by
MAGBOLTZ~\cite{magboltz} calculations, gives $A \rho =(3160\pm 90)$ cm$^{-1}$ and $B \rho =(136.4\pm 1.0)~$kV/cm
for pure argon at 87~K and 1~bar.

For amplification in holes, it is convenient to make
use of the same formalism by replacing the gap thickness $d$
by the effective amplification path length within the holes, called $x$, which can be related
to the length of the field plateau along the hole.
The effective gain is then expressed as:
\begin{equation}
G_{eff} \equiv  T \exp\left[A \rho x e^{(-B \rho/( \kappa E_0 )})\right]
\label{eq:gain}
\end{equation}
where $A\rho$, $B\rho$ are given by MAGBOLTZ, $E_0=V/d$ is the nominal electric field
and $d$ the LEM thickness, $\kappa E_0$ the effective field,
and $T$ is the transparency
(some electrons, in part due to the large diffusion in pure argon, may be collected on the grids or on the electrodes of the LEM).
Electrostatic calculations of the LEM-hole geometry give a maximum field in the hole which is lower than  the naive $V/d$, consistent
with  a value of $\kappa = 0.95$ and an effective length in the range of $0.7$~mm for a 1~mm thick LEM.

Electron multiplication
in holes has been experimentally investigated for a large number of applications.
The most extensively studied device is the  Gas Electron Multiplier~(GEM)~\cite{Sauli:1997qp}, 
made of
50--70~$\mu$m diameter holes etched in a 50~$\mu$m thick metallized
Kapton foil.  Stable operation has been shown with various gas mixtures
and very high gains.
An important step was the operation of the GEM in 
pure Ar at normal pressure and temperature~\cite{Sauli99}. 
Rather high gains were obtained, of the order of 1000, supporting  evidence
for the avalanche  confinement to the GEM micro-holes.
Operation of GEMs in an avalanche mode in pure Ar in double phase conditions has been
studied~\cite{Bondar06}, 
using triple-stage GEMs reaching gains of the order of 5000.

The successes of the GEMs triggered the concept of the Large Electron
Multiplier~(LEM) also called thick GEM (THGEM) (for a review see~\cite{Bondar08}), 
a coarser but more rigid structure made with sub-millimeter-size holes in a millimeter-thick
printed circuit board~(PCB).
Compared to GEMs, LEMs are mechanically more robust, are sturdy to withstand cryogenic temperatures and stresses,
and have a good discharge hardness. 

\begin{figure}[htbp]
   	\centering
   	\includegraphics [width=0.80\textwidth]{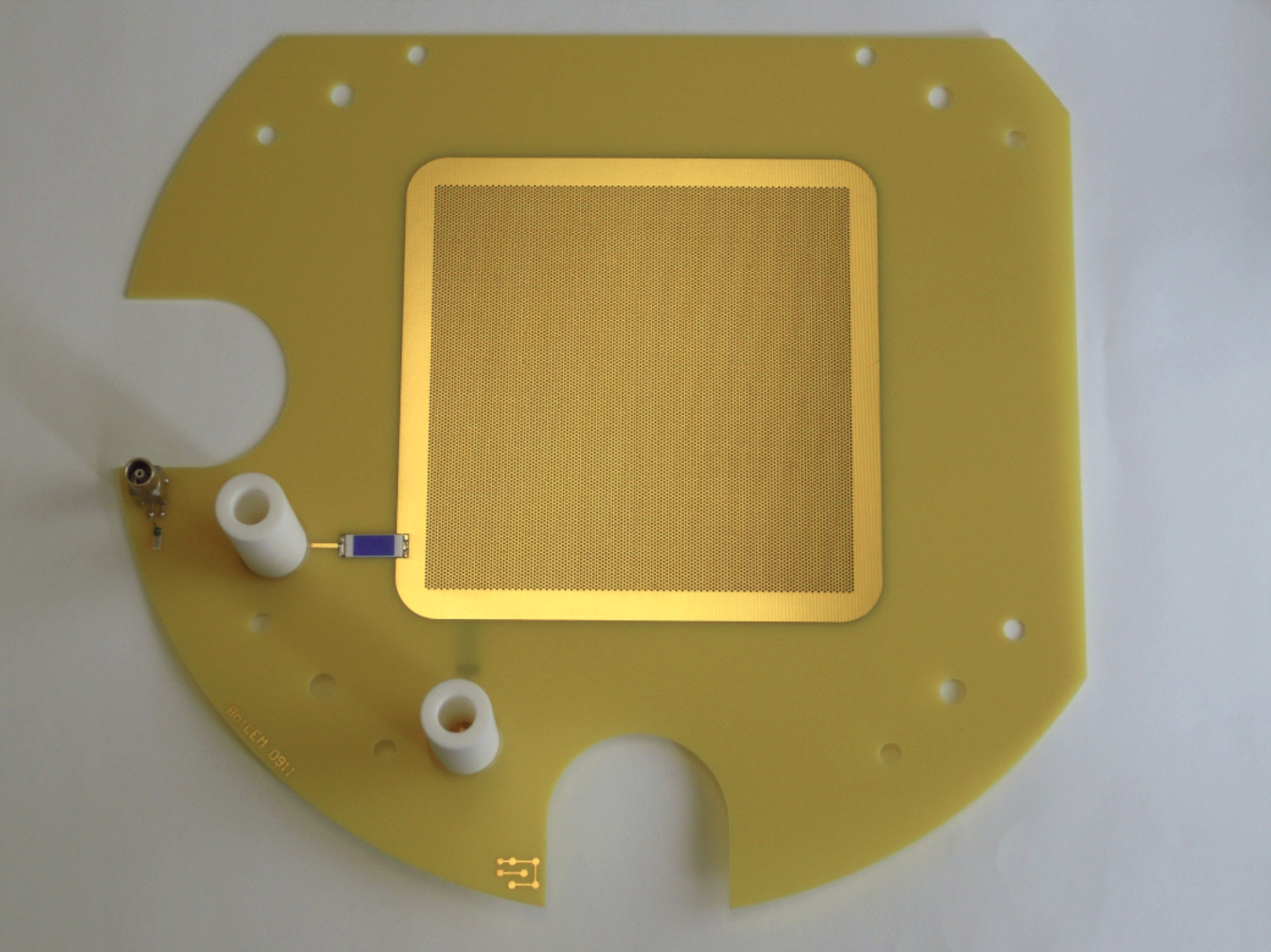} 
   	\caption{Picture of a 1 mm thick LEM with two Macor insulated HV connections and a 500~M$\Omega$ resistor.}
   	\label{fig:LEM}
\end{figure}

\begin{figure}[htbp]
   	\centering
   	\includegraphics [width=0.80\textwidth]{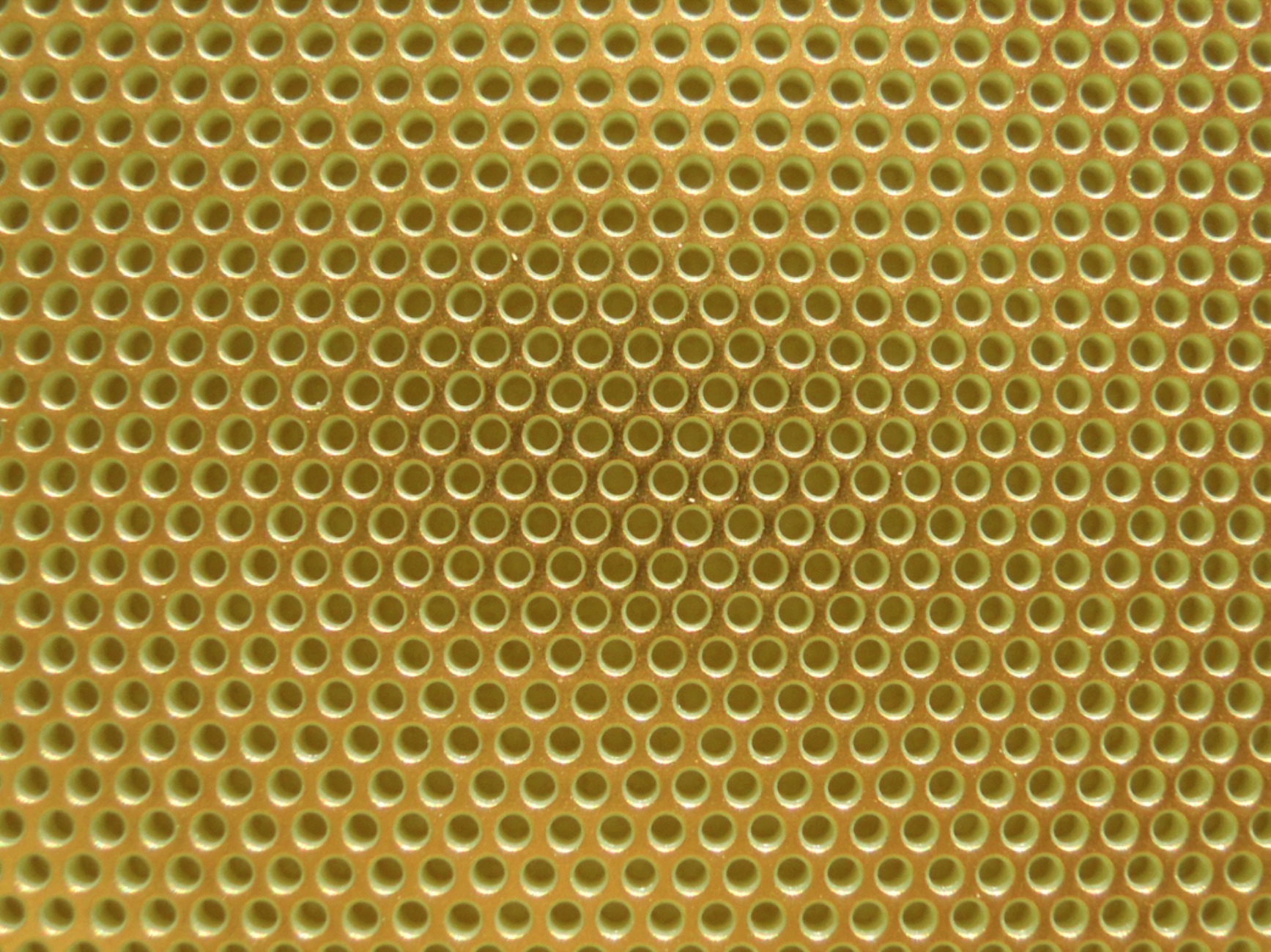}
   	\caption{Close up of the LEM showing the well centered and uniform dielectric rims.}
   	\label{fig:LEMclose}
\end{figure}

We have built several LEM prototypes using standard PCB techniques from different manufacturers.
Double-sided copper-clad (of 16-50~$\mu$m layers) FR4 plates with thicknesses ranging from 0.8~mm to 1.6~mm
were drilled with a regular pattern of 500~$\mu$m diameter holes  at a relative distance of 800~$\mu$m. 
Table~\ref{tab:LEMDesign} shows the parameters of the 1~mm thick LEMs used for this test and
a picture of one such LEM is shown in Figure~\ref{fig:LEM}.
The LEMs were manufactured at the CERN TS/DEM workshop with a specific technique to ensure alignment between holes and rims around
the holes: the holes are Computer Numerical Control (CNC) drilled and dielectric rims are etched uniformly around the holes~\cite{oliveira}.
The copper layer thickness before etching was about 50~$\mu$m and it reduced to about 30~$\mu$m after rim etching.
The resulting thickness of the copper layer is sufficient in order to have resistance to sparks.
Thanks to this manufacturing technique,
the dielectric rims are perfectly aligned with the holes (see Figure~\ref{fig:LEMclose}).
Since the dielectric rim reduces the probability of a discharge~\cite{Breskin:2008cb} higher voltages can be applied between the 
two LEM electrodes. This increases the maximum gain that can be reached before a breakdown occurs. 

\begin{table}[htbp]
	\centering
	\begin{tabular}{lr} 
       		\hline        
     		\hline        
       		size				&  10x10 cm$^2$	\\
		PCB thickness		&  1 mm			\\
		copper layer thickness (before etching) &  50 $\mu$m		\\
		copper layer thickness (after etching) &  $\approx$ 30 $\mu$m		\\
        		hole diameter		&  500 $\mu$m		\\
        		hole pitch			&  800 $\mu$m		\\
        		dielectric rim size	&  50-60 $\mu$m	\\
       		\hline        
     		\hline        
	\end{tabular}
	\caption{Physical parameters of the Large Electron Multiplier (LEM).}
	\label{tab:LEMDesign}
\end{table}

\section{The projective 2D anode}
\label{sec:anode}
\begin{figure}[htbp]
   	\centering
   	\includegraphics [width=1.0\textwidth]{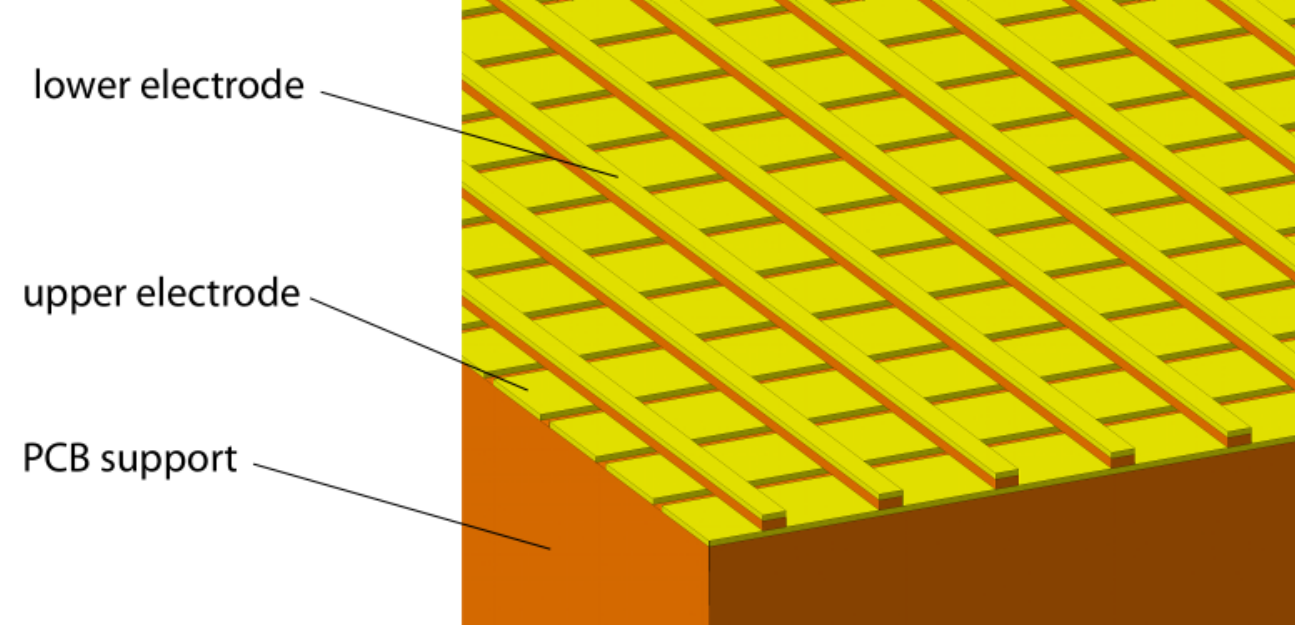}
   	\caption{Illustration of the projective 2D anode. See text.}
	\label{fig:proj2danode}
\end{figure}

The 2D anode provides X- and Y-views, i.e. two independent planes with mm-scale readout pitch. 
It is based on the GEM 2D readout concept~\cite{Bressan:1998jj} and
is conceptually shown in Figure~\ref{fig:proj2danode}. It is composed 
of two independent sets of strips oriented perpendicularly to each other, called ``lower'' and ``upper'' electrodes, and separated by a 
thin insulating film. Both sets of electrodes are operated at the same electric potential,
and the geometry of their strips is chosen such that each electrode collects half of the charge.
We note that
the strips of the upper electrode are wider to compensate the screening of the thin strips of the lower electrode.

\begin{figure}[htbp]
   	\centering
   	\includegraphics [width=0.7\textwidth]{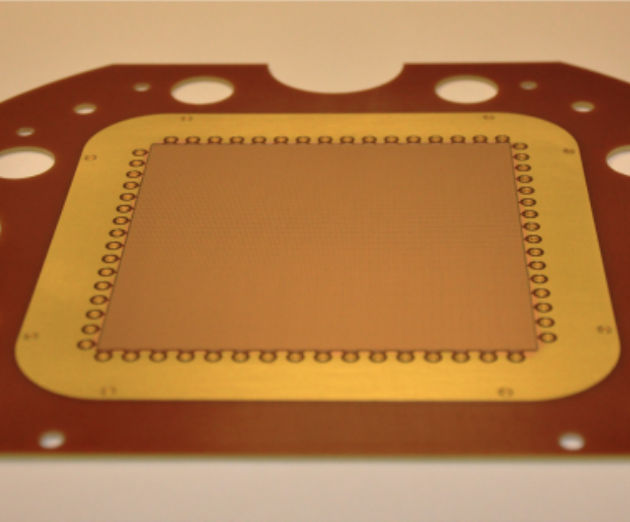}
   	\includegraphics [width=0.7\textwidth]{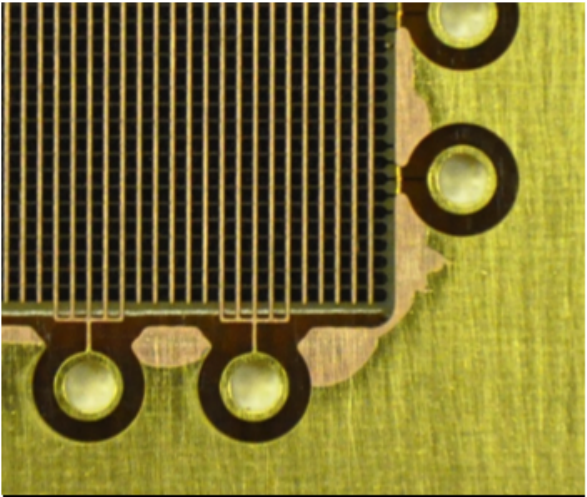}
   	\caption{Pictures of the manufactured projective 2D anode prototype. (top) overall view (the readout area is 10x10~cm$^2$). (bottom) zoom on the strips
	showing the bridging of 5 strips to obtain a 3~mm readout pitch.}
	\label{fig:anode}
\end{figure}

A prototype projective anode used for this test with the parameters described in Table~\ref{tab:anodeDesign}
was produced by the CERN TS/DEM workshop, using photolithographic etching techniques.
Pictures are shown in Figure~\ref{fig:anode}. To produce the anode,
one of the two copper surfaces of an insulating 50~$\mu$m thick polyimide foil was etched with the pattern
of the strips of the upper electrode and glued on a PCB support. The other copper surface
was masked with the pattern of the lower electrode, first the copper and afterwards the polyimide
were etched away leaving the strips of the upper electrode exposed.
A fine strip pitch of 600~$\mu$m ensures that charge clouds always induce signals on more than one strip
of the electrodes. 
The final readout pitch of 3 mm is reached by electrically bridging 5 strips, as can be seen in 
Figure~\ref{fig:anode}(bottom).

\begin{table}[h]
   	\centering
   	\begin{tabular}{lr} 
       		\hline
		\hline
		readout pitch    	&  3 mm 			\\
       		strip pitch 			&  600 $\mu$m 	\\
       		lower electrode width  	&  120 $\mu$m 	\\
       		upper electrode width  	&  500 $\mu$m 	\\
     		kapton thickness   	&  50 $\mu$m 		\\
       		\hline
		\hline
	\end{tabular}
	\caption{Design parameters of the projective 2D anode.}
   	\label{tab:anodeDesign}
\end{table}

\section{Operation in double phase liquid Argon}
\label{sec:operation}
In order to be able to measure tracks in the whole active volume, it
is necessary that the mean lifetime of the drifting free electrons
in LAr is at least of the order of the maximal drift time. 
A drift of 21~cm with an electric field of 500 V/cm without significant charge loss requires the LAr 
to contain less than few ppb of electronegative impurities like O$_2$, H$_2$O and CO$_2$. 
This makes it necessary to purify the commercial grade
LAr (declared content of $O_2 < 5$ ppm and $H_2O < 10$ ppm) used to
fill the detector. 

Precautions are used to ensure a good purity of the argon during operation:
after the assembly, the detector, the recirculation system and
the cartridges are pumped down to $<10^{-6}$~mbar. To speed up the
cleaning we warm the detector vessel from outside without exceeding
50$^o$~C, mainly not to damage the photocathode of the PMT.

An external cooling bath is filled with LAr in order to cool the detector down to 87 K. 
After it reaches the final temperature, the inner vessel is filled with argon.
During the filling, argon is fed into a custom-made filter,
filled with pure copper powder. The copper oxidizes trapping oxygen
impurities from the liquid; this cartridge can be regenerated by
increasing the temperature and flushing argon-hydrogen mixture to
induce CuO$_2$+2H$_2$$\rightarrow$Cu+2H$_2$O.

To maintain the purity for long periods the liquid argon is evaporated
and pushed by a metal bellows pump through a commercial
SAES getter\footnote{SAES Pure Gas Inc., MicroTorr MC400.}.
(See Ref.~\cite{Badertscher:2009av} for details on the purification system). 
The cleaned gas argon is re-condensed into the detector by means of a
spiral condenser. The argon flow is regulated via a needle valve at
the input of the pump and it is monitored with a mass flow meter
mounted between the pump and the getter.
We operate the recirculation at 7.5~slm: two days are needed for a
full volume change.

Moreover the argon filling procedure is
crucial: the recirculation system is turned
on while cooling the detector. This procedure has two advantages: the
outgassed molecules are trapped into the getter cartridge during the
recirculation and the presence of argon gas inside the detector favors
uniform and fast cooling of all the the parts.

The free electron lifetime measurement which was done about one day after the filling indicated an initial purity of about 1~ppb. 
The corresponding plot with the description of the analysis is shown in the next section. After the filling the detector has been 
operated during one week without any argon purification. 

\begin{figure}[h]
   	\centering
   	\includegraphics [width=0.95\textwidth]{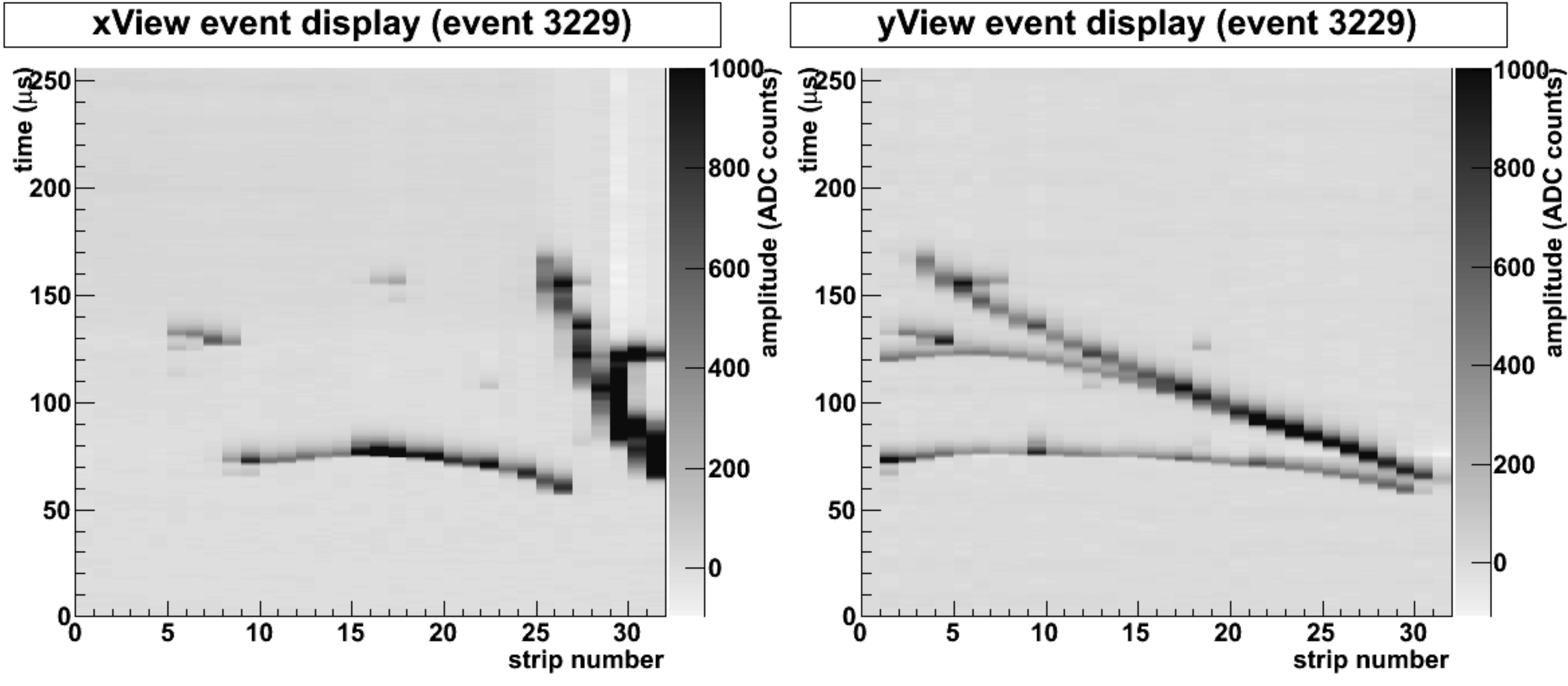}
   	\includegraphics [width=0.95\textwidth]{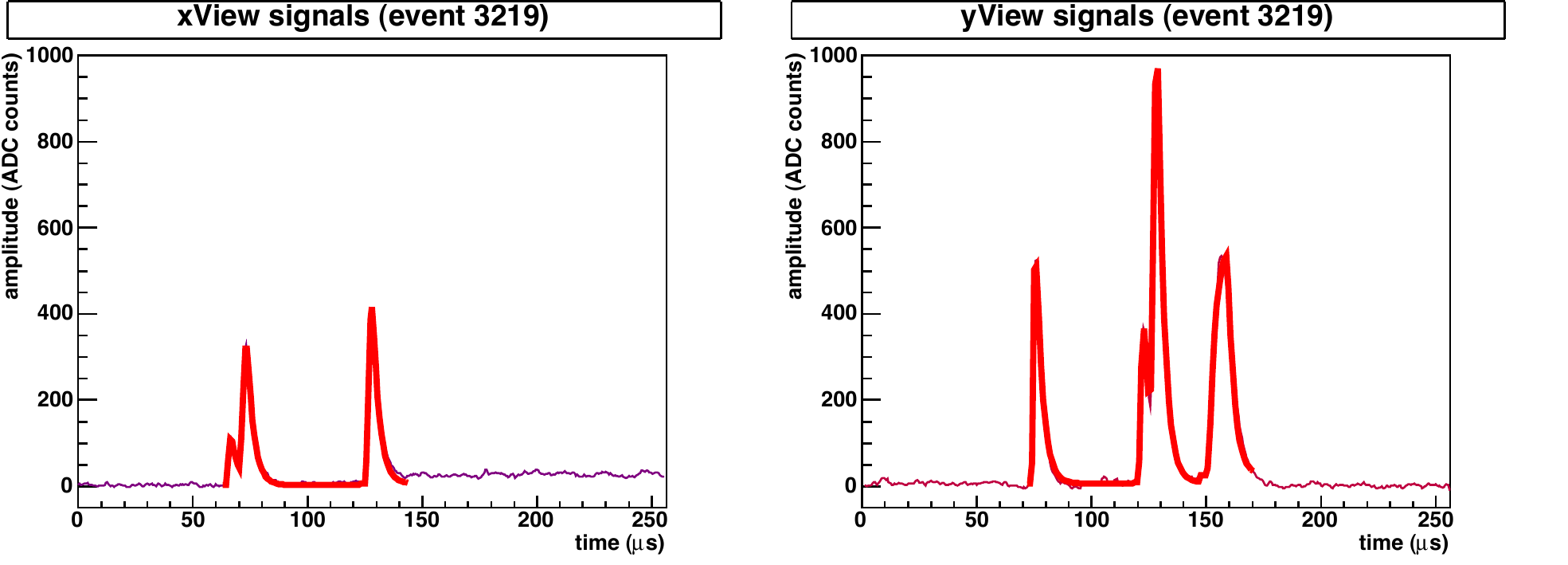}
   	\caption{Multi-track event. \emph{Top}: event display: channel number vs. drift time, x-view \emph{left}, and y-view \emph{right}. The grey scale is proportional to the signal amplitude. \emph{Bottom:} Typical waveforms for the x-view \emph{left} and the y-view \emph{right}. The thick red curve shows the result
	of the signal fit using the response of the preamplifier.}
   	\label{fig:event_gall}
\end{figure}

\begin{figure}[h]
   	\centering
   	\includegraphics [width=0.95\textwidth]{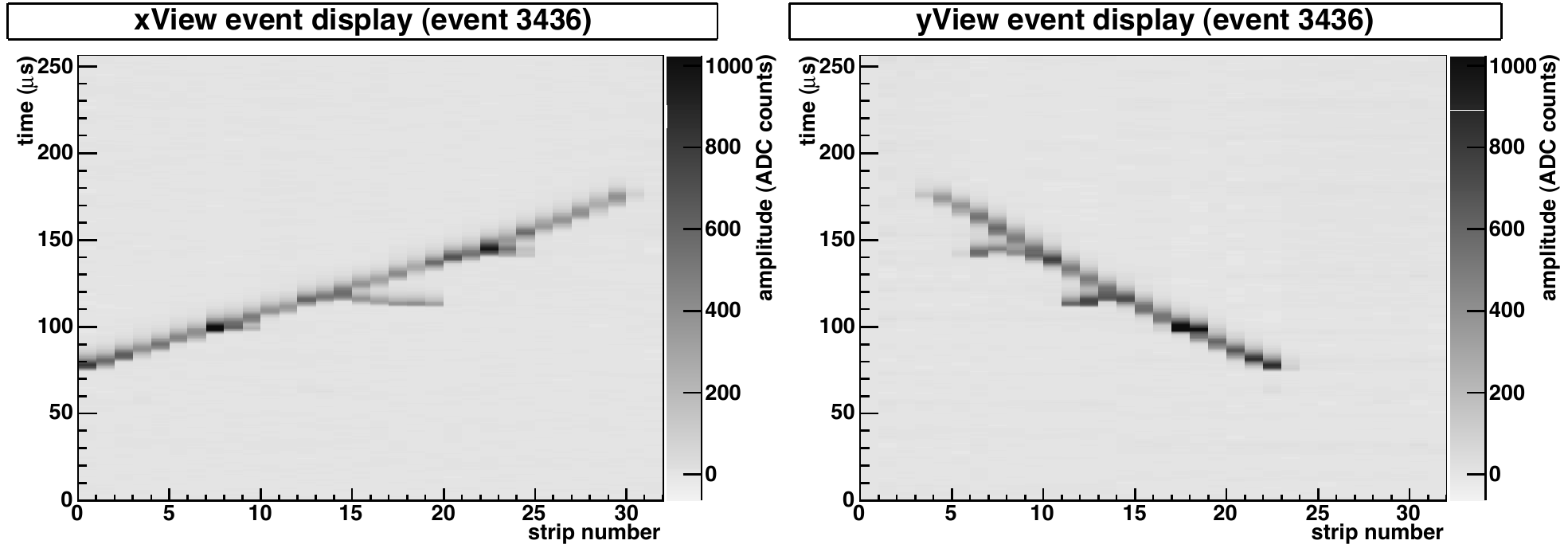}
   	\includegraphics [width=0.95\textwidth]{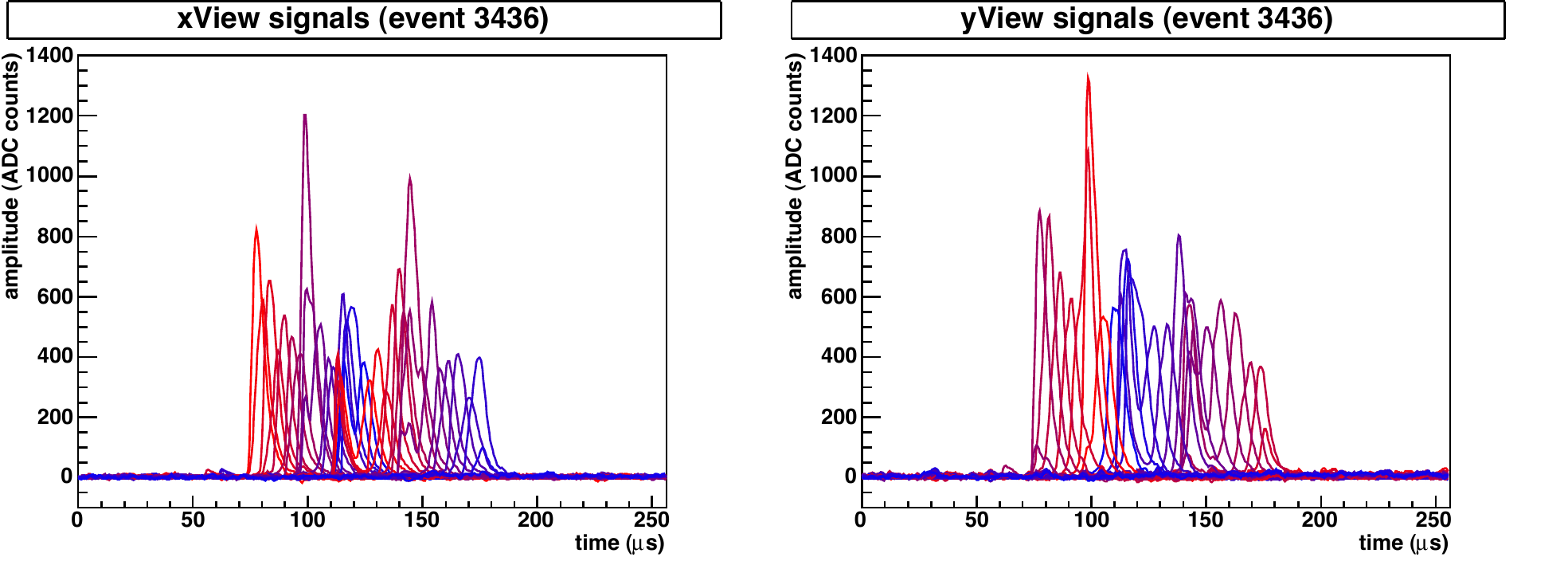}
   	\caption{Muon crossing the chamber with visible delta rays. \emph{Top}: event display: channel number vs. drift time, x-view \emph{left}, and y-view \emph{right}. The grey scale is proportional to the signal amplitude. \emph{Bottom:} 32 waveforms of the x-view \emph{left} and the y-view \emph{right}.}
   	\label{fig:event1}
\end{figure}

\begin{figure}[h]
   	\centering
   	\includegraphics [width=0.95\textwidth]{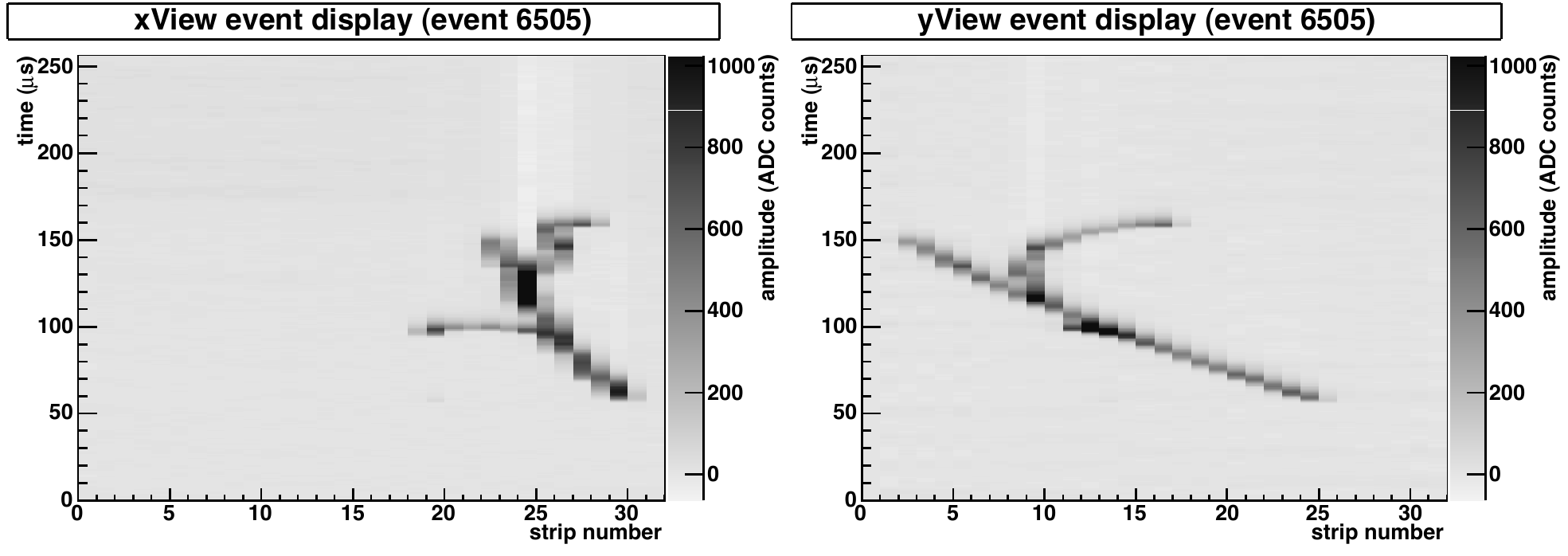}
   	\includegraphics [width=0.95\textwidth]{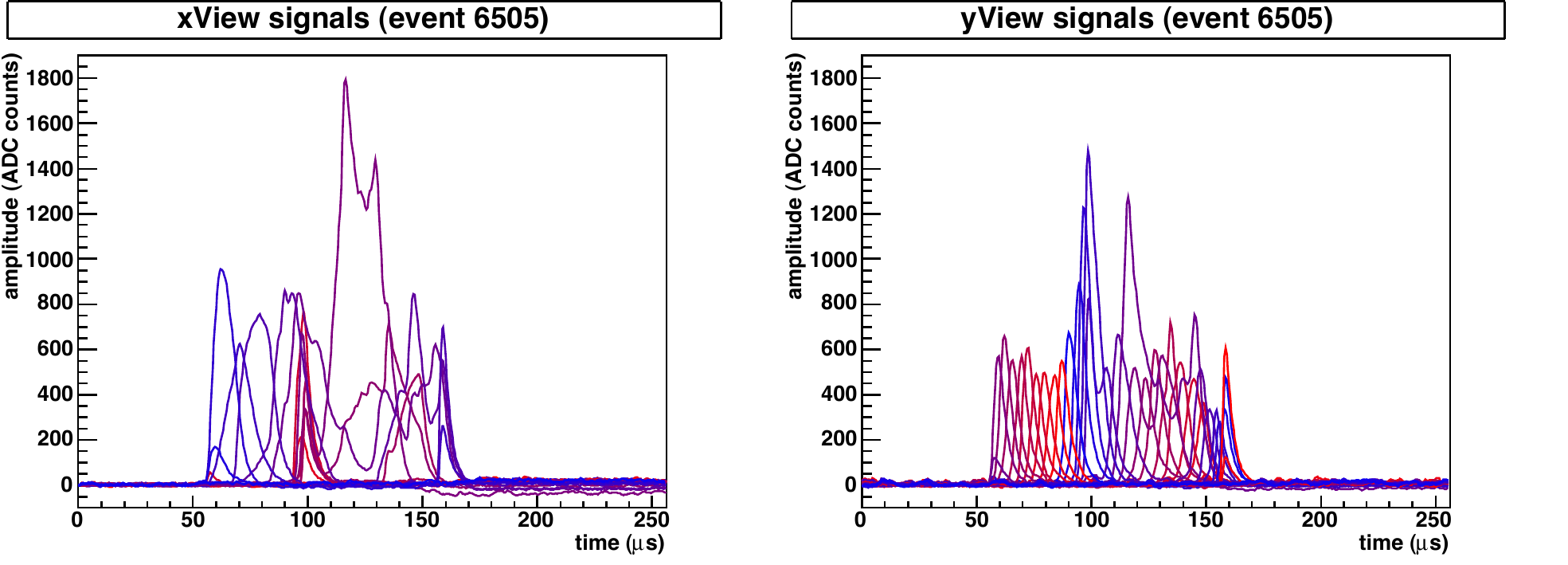}
   	\caption{ibid, see caption of Fig. \ref{fig:event1}}
   	\label{fig:event2}
\end{figure}

\section{Muon track reconstruction and free electron lifetime correction}
\label{sec:reco}
The aim of this section is to describe the reconstruction procedure which was used to select and evaluate long tracks. 

Because the ground configuration of this setup was not fully optimized, coherent noise was observed on all channels. 
After applying a mean trace subtraction of channels without a signal 
to remove such coherent noise, waveforms are very clean. Figure~\ref{fig:event_gall} shows a collected multi-track event
after the digital filter.
Hits are discriminated from the residual noise by a hit finding algorithm. 
Found hits fit the well-known response function of the preamplifier convoluted 
with a constant current at its input, whose length is left as a free parameter. 
Typical fit results are shown with thick curves on the waveform plots.

After a calibration with charge pulses the signal integral 
is converted into the charge collected on the strip. 
Besides the charge of a hit, the start time of the signal is also extracted from the fit. The drift coordinate 
$z$ of the hit is then given by the electron drift velocity multiplied by 
the time difference between the trigger signal from the PMT and the start time of the signal. The strip number gives 
the coordinate of the corresponding view. 

Figure~\ref{fig:event1} and~\ref{fig:event2} show two ionizing events with long tracks, 
triggered by the primary scintillation light which was detected with the PMT. 
After clustering of connected hits, matching 2D tracks were fitted with the linear functions $z(x)$ and $z(y)$ which could then be solved to extract the missing coordinates (y of X-view and x of Y-view). Eventually long tracks crossing the full active volume were selected based on cuts on the 
endpoints of the track and also the goodness of the linear fits. 

\begin{figure}[tbp]
   	\centering
    	\includegraphics [scale=0.5]{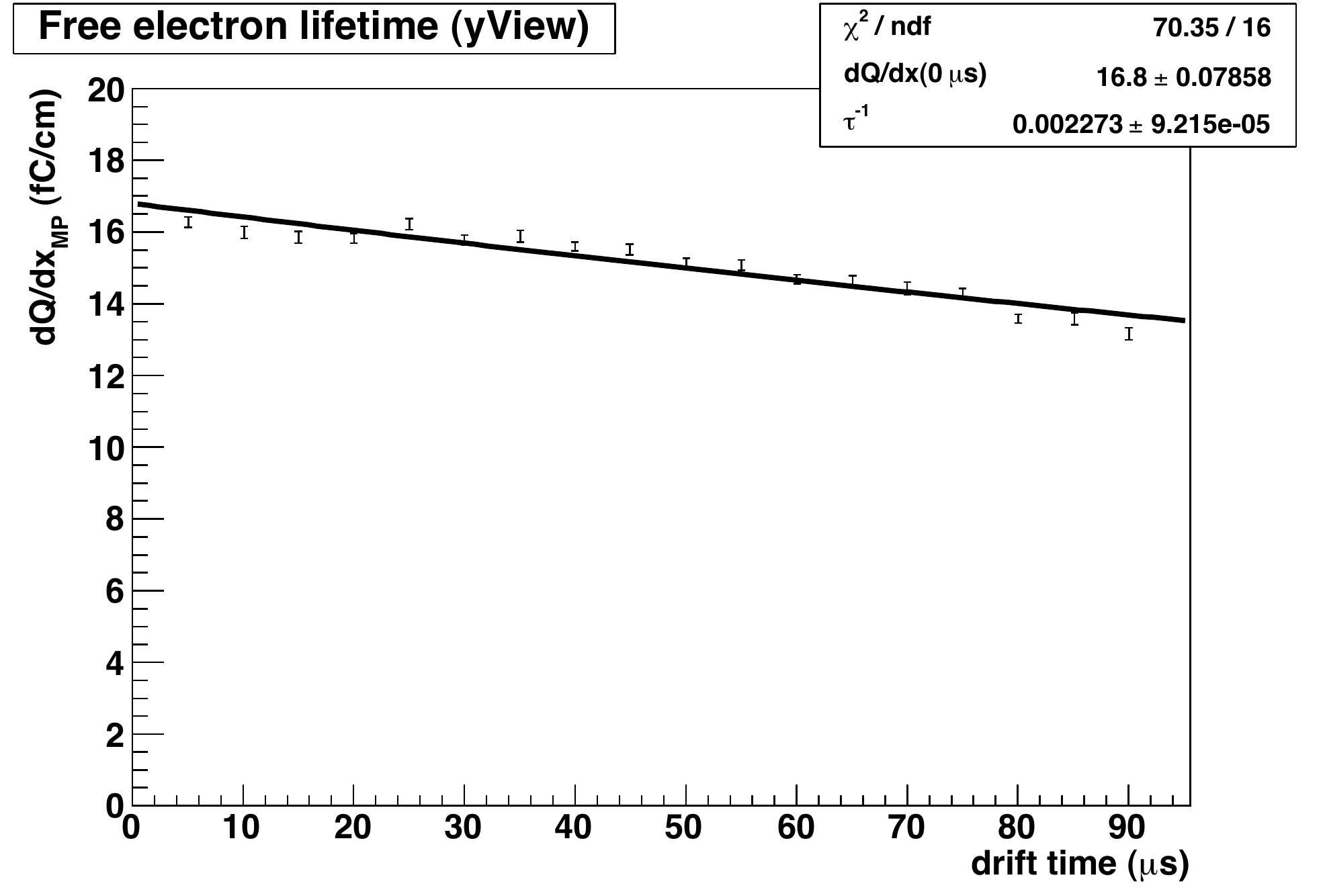}
   	\caption{Most-probable stopping power ($dQ/dx)_{\mathrm MP}$ in fC/cm for different slices of drift time. 
	The curve is an exponential fit characterizing the free electron lifetime.
	The error bars show the statistical error of the Landau fit (see text). For this data, the LEM gain was about 3.}
   	\label{fig:lifetime}
\end{figure}

The energy loss of the long muon tracks was used to estimate the effective gain of the chamber. The $dQ/dx$ value of each hit is defined by the ratio of the hit integral and the track length below the considered strip. After dividing the active volume into slices of equal drift length,
a Gauss convoluted Landau function was fitted to each of the obtained collected charge per unit length distributions.
By plotting the most-probable values of these distributions
against the drift time, the free electron lifetime can be obtained by fitting an exponential decay to the points. See Figure~\ref{fig:lifetime}. The error bars show the statistical error of the Gauss-convoluted Landau fit. The measured free electron lifetime is ($440\pm20$)~$\mu$s, which corresponds to a concentration of about ($0.68\pm0.03$)~ppb of oxygen equivalent impurities in LAr~\cite{Buckley:1988qx}. 

\section{Signal amplification and effective gain}
\label{sec:results}
\begin{figure}[htbp]
   	\centering
     	\includegraphics [scale=0.325]{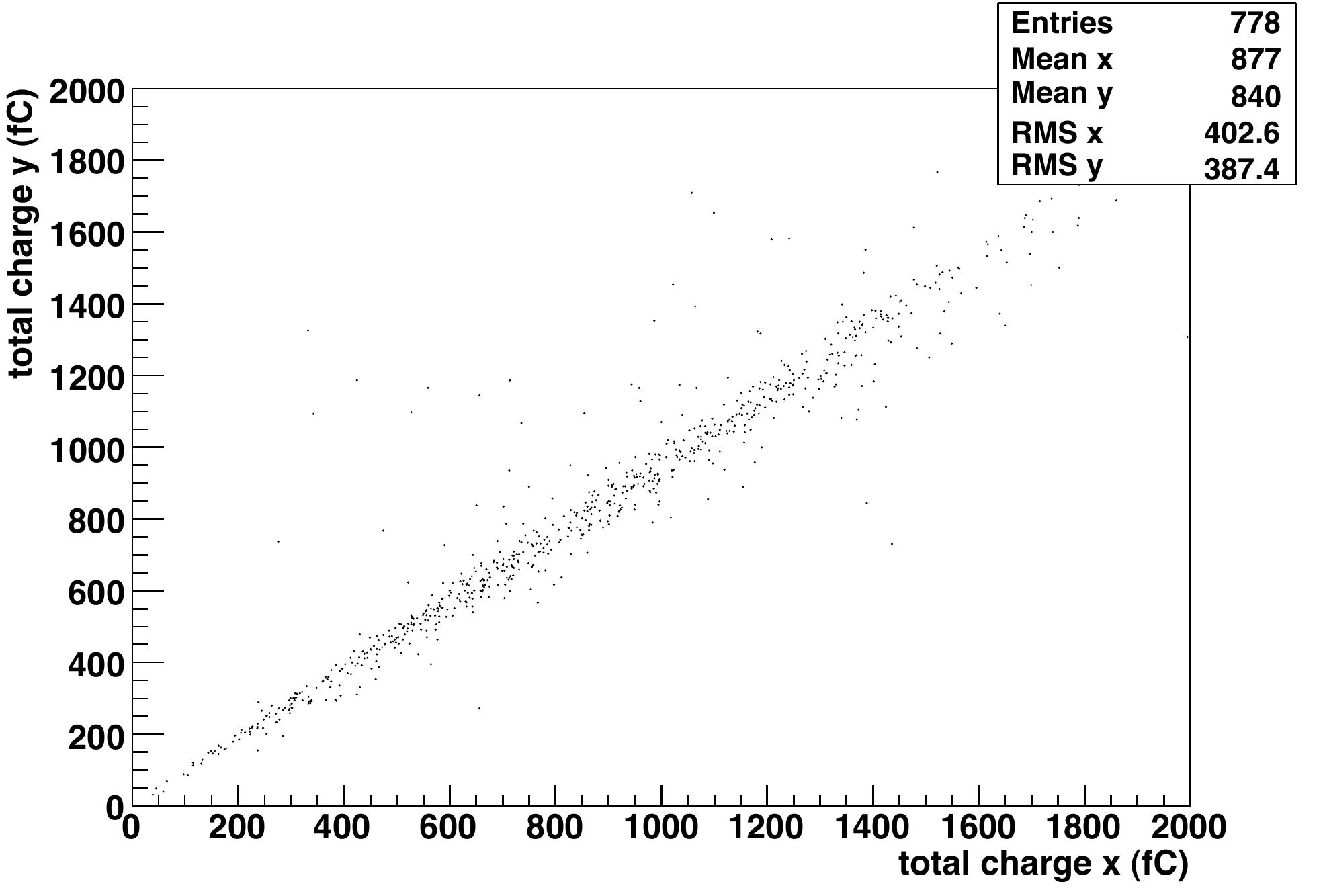}
     	\includegraphics [scale=0.325]{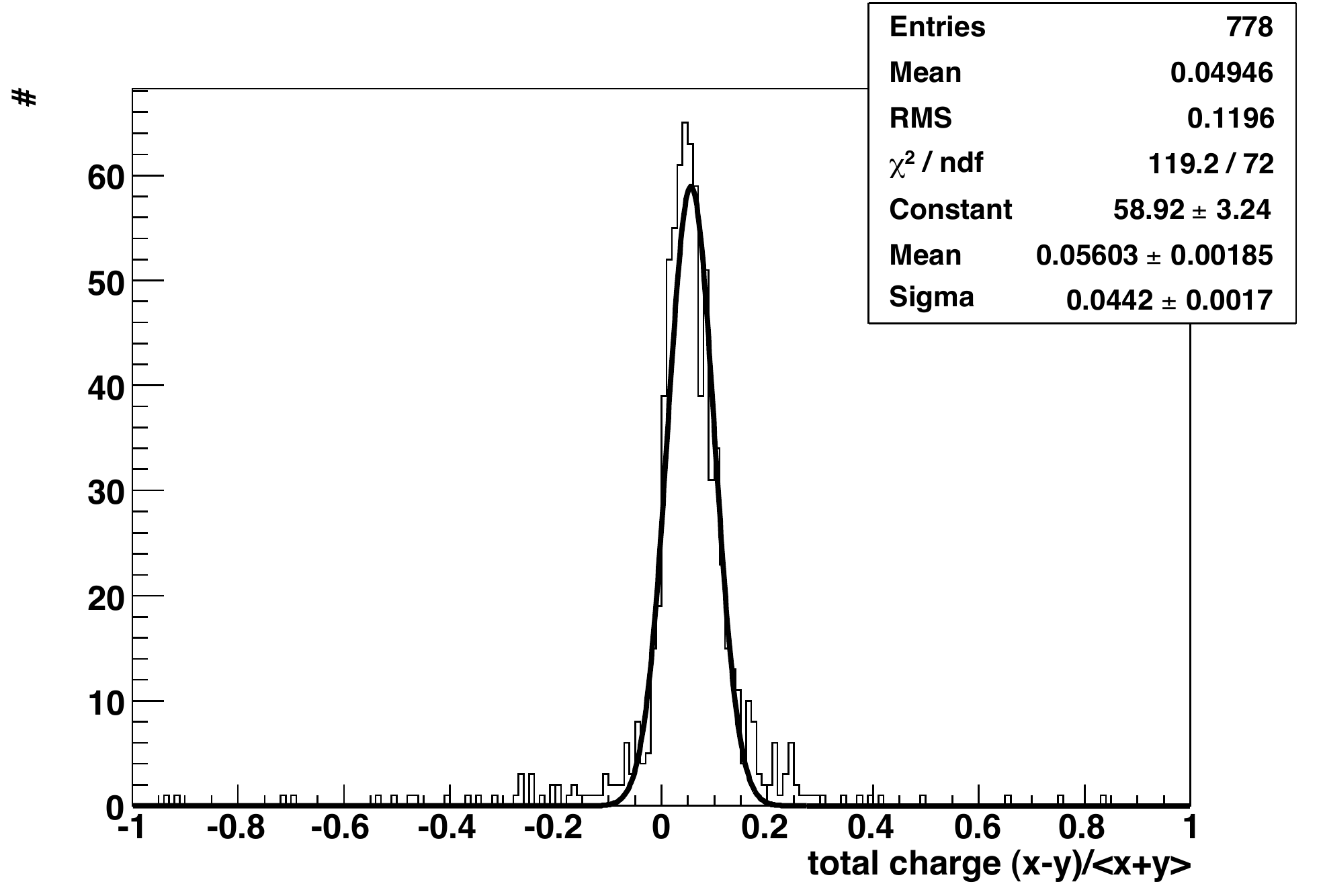}
   	\caption{Left: Scatter plot of total charge collected on the x-view versus the charge collected on the y-view. Right: histogram of the charge collection difference, normalized by the average charge.}
   	\label{fig:sharing}
\end{figure}

The signal induction on the two views is based on the principle of charge sharing between them. In order to get a quantitative measurement, long muon tracks, selected as described in the previous section, were used. The left plot of Figure~\ref{fig:sharing} shows a scatter plot of the total amount of charge collected on the X- vs the Y-view. As shown in the histogram on the right side of the Figure the normalized difference of the charge collected on the two views is better than~5\%. 

The second issue which has been addressed is the amplification of the system.
The {\it effective} gain is defined as the sum of the collected charge per unit length on the two anode views (corrected for the drifting electron lifetime) divided by the charge per unit length released in liquid argon by a minimum ionizing muon at the given electric field. The former is estimated from the $dQ/dx$ distribution of the reconstructed muon tracks.
Minimum ionizing particles (mip) deposit in LAr an energy of 1.519~MeV cm$^2/$g. 
For our drift field configuration of 500 V/cm and taking into account 30\% recombination, a total charge
of 10~fC/cm is released for mip tracks. 

The effective gain, in addition to the charge multiplication in the LEM holes, takes into account potential charge losses from the efficiency to extract the electrons to the vapor phase and from the transparency of the grids and the LEM. 
For this reason the effective gain is sensitive to all the details of the electric field configuration. 
The working point is found from electrostatic computations of the drift lines (see Section~\ref{sec:driftfield}) and from the optimization of the transparency. 

\begin{figure}[htbp]
   	\centering
   	\includegraphics [width=0.9\textwidth]{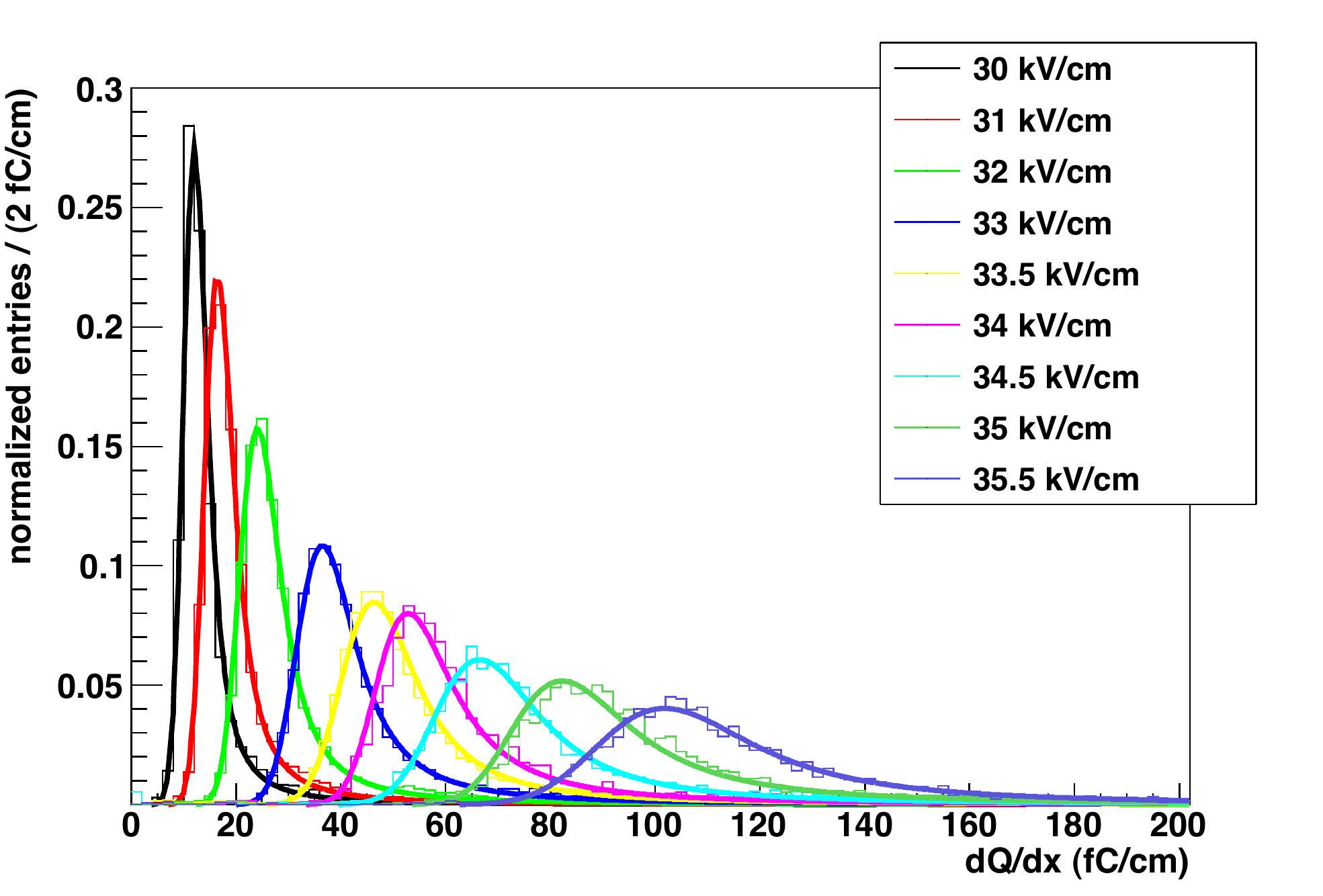} 
   	\caption{$dQ/dx$ distributions of reconstructed cosmic muon tracks with different LEM nominal fields.
	 A Gauss convoluted Landau Function is fitted to each distribution.}
   	\label{fig:dQdx}
\end{figure}

\begin{figure}[htbp]
  	\centering
   	\includegraphics [width=0.9\textwidth]{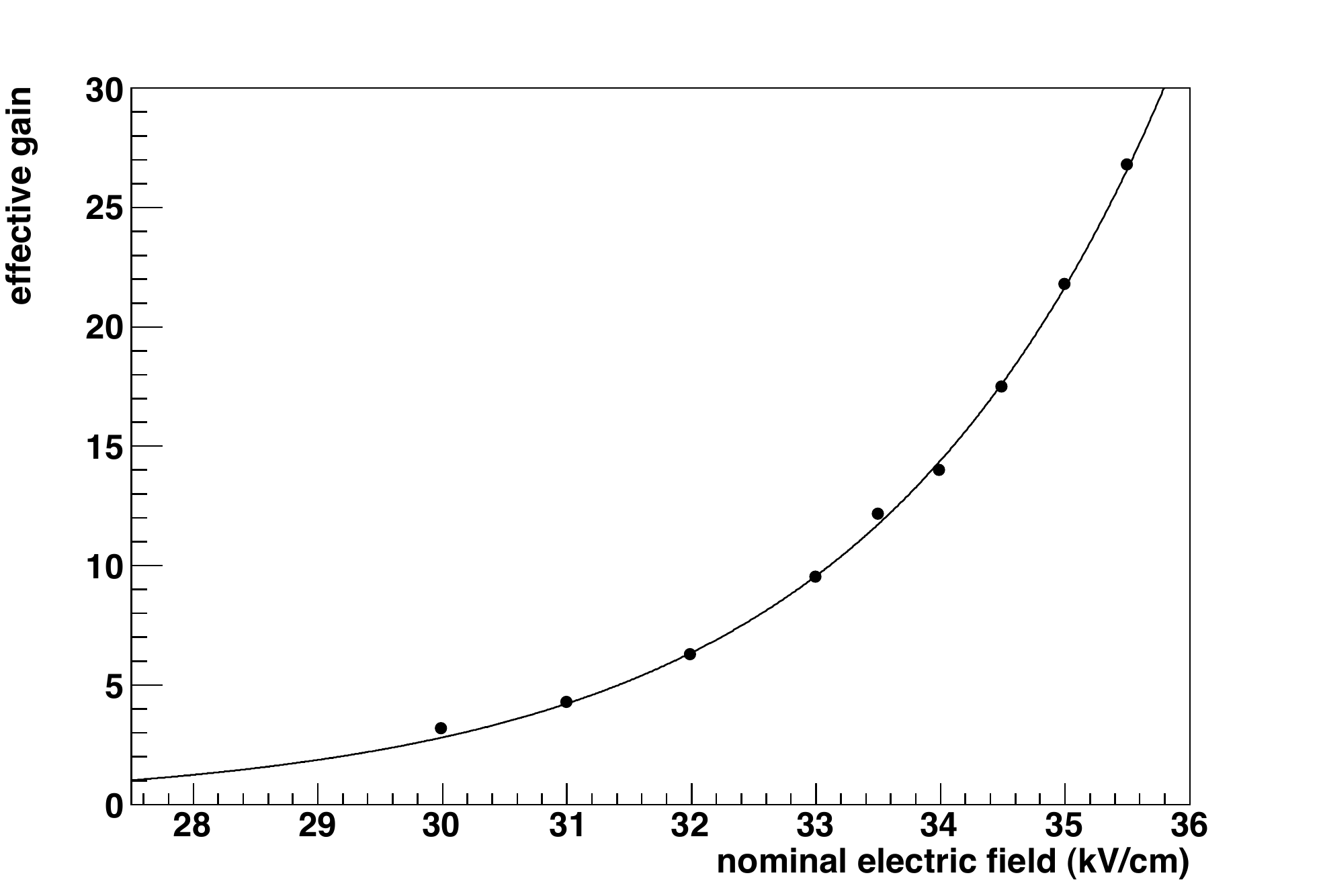}
   	\caption{Effective gain as a function of the nominal electric field in the LEM holes. 
	The curve is a model fit to the measured gain. See text.}
   	\label{fig:gain}
\end{figure}

In order to measure the gain as a function of the amplification field the voltage across the LEM electrodes was increased from 3000 V up to 3550 V, where first discharges induced by ionizing events occurred, while leaving the other fields constant. Figure~\ref{fig:dQdx} shows the normalized $dQ/dx$ distributions of long cosmic muon tracks measured with one of the two views. All the distributions were fitted with a Gauss convoluted Landau function. Figure~\ref{fig:gain} shows the effective gain as a function of the applied voltage.  

\begin{figure}[htb]
  	\centering
   	\includegraphics [width=0.95\textwidth]{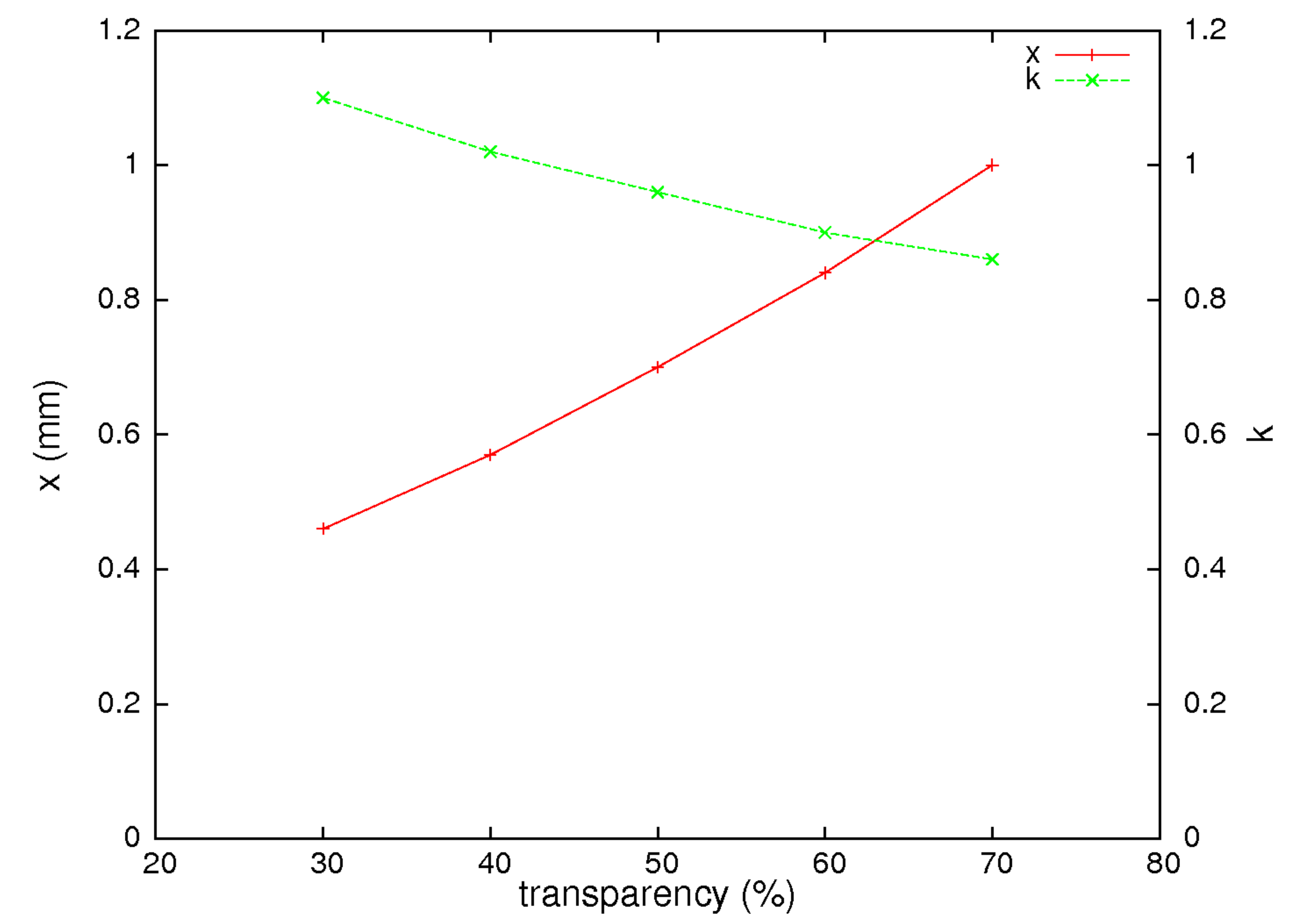}
   	\caption{Fitted effective thickness $x$ and the effective field parameter $\kappa$
	as a function of transparency $T$. See text.}
   	\label{fig:fitxk}
\end{figure}

The maximum effective gain obtained in stable conditions is 27, with a signal to noise ratio larger than 200 for minimum ionizing particles. 
The measured effective gains as a function of the applied voltage can be compared to the predictions,
as presented in Section~\ref{sec:LEM}. The gain curve fits the functional form in Eq.~\ref{eq:gain} and one
determines the effective thickness $x$ and the effective field parameter $\kappa$. Figure~\ref{fig:fitxk} shows
the fitted values as a function of the  assumed transparency $T$. It demonstrates that 
the observed gain curve is consistent with expectations. If we assume a value $\sim 50$\% for the 
transparency, the data point to $x=0.7$~mm, and $\kappa=0.96$. These values are perfectly 
compatible with the expectations from electrostatic field calculations.

\begin{figure}[htb]
   	\centering
   	\includegraphics [width=0.9\textwidth]{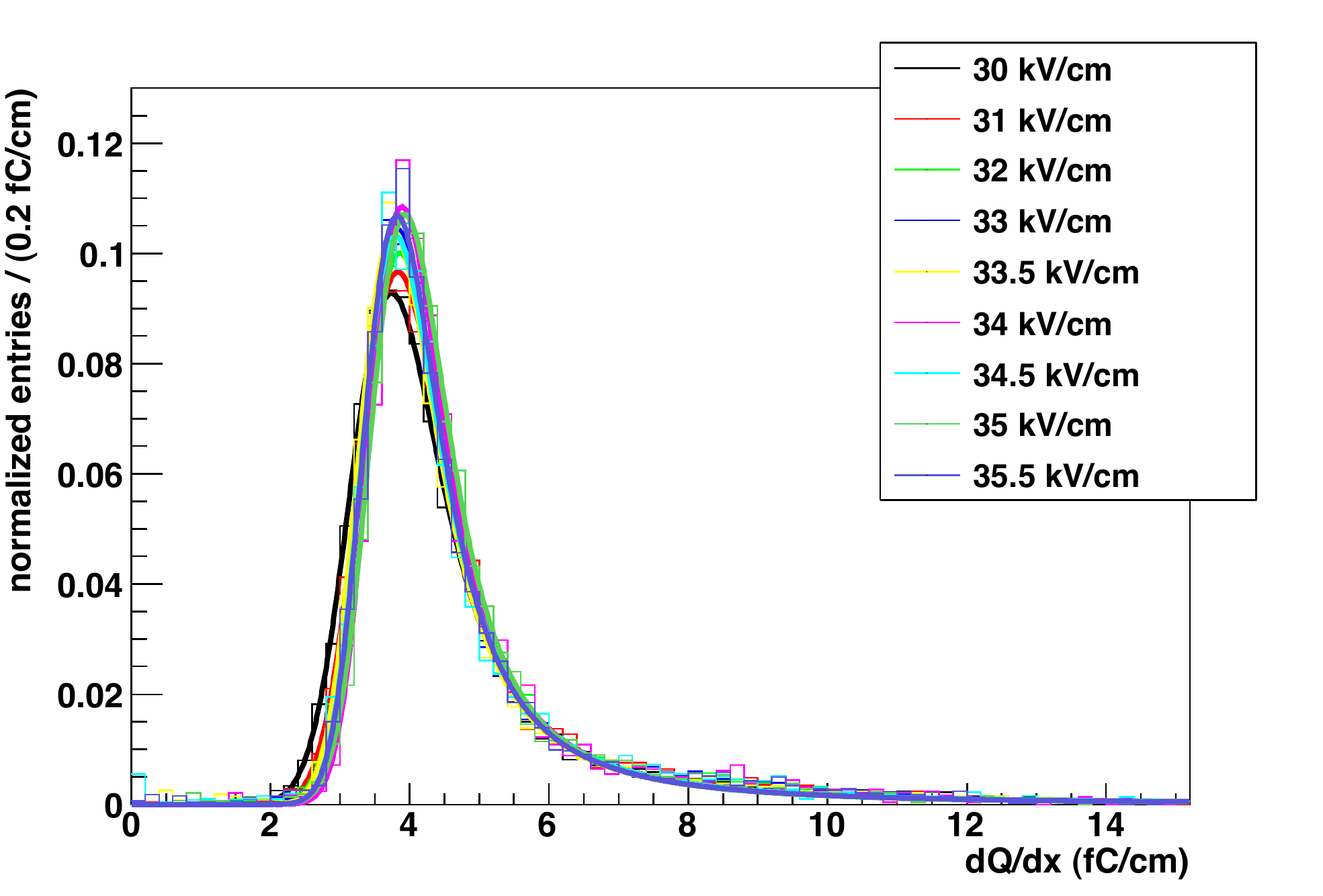} 
   	\caption{Distribution of the gain normalized $dQ/dx/G$ (fC/cm) of reconstructed tracks, for different LEM nominal fields. 
	In order to compare the distribution measured at different gains, we divided the $dQ/dx$ entries by the expected gain.
	The	measured width of the Landau distribution, characteristic of the medium, is unaffected by the gain.}
   	\label{fig:dQdxnorm}
\end{figure}

We can also address the charge measurement resolution and its potential degradation due to fluctuations of the amplification in the holes.
In order to compare the different situations, we study the $(dQ/dx)/G$ distributions, i.e. the measured $dQ/dx$ divided by the gain~G,
obtained with the reconstructed cosmic muon tracks.
The histograms, with the corresponding fits, for different LEM fields are overlapped in Figure~\ref{fig:dQdxnorm}. 
The widths, characterized by the
Landau distribution
of the medium and the resolution, are similar for all gains, showing that the amplification stage does not
degrade the observed $dE/dx$ distribution.

\section{Conclusion}
\label{sec:conclusion}

The LAr LEM-TPC is a novel kind of complete calorimetric and tracking detector capable of charge multiplication.
For the first time we operated a detector based on a
decoupled LEM amplification stage, that allows adjustable charge amplification in
argon vapor, and a 2D projective readout.

From the collected set of cosmic muon tracks, we observe the considerable improvement in
the signal to noise ratio obtained thanks to the gain, opening the path to very long drift liquid Argon TPCs.
The achieved stable gain of 27 can compensate for e.g. the attenuation of the collected
charge for drifts of the order of 10~m even in the presence of diffusion or 
charge attenuation from impurities.
For direct Dark Matter search with imaging, higher gains ($\gtrapprox$~100) are
needed to reach an energy threshold down to tens of keV for nuclear
recoil detection.

This proof of principle therefore represents another important milestone in the realization of very large, long drift (cost-effective) LAr detectors for next generation neutrino physics and proton decay experiments, as well as for direct search of Dark Matter with imaging devices.
We believe that this technology is scalable up to very large detectors, for instance
instrumenting the surface with a collection of $\sim$1$\times$1~m$^2$
independently operating units.

\section*{Acknowledgements}
This work was supported by ETH Z\"urich and the Swiss National Science Foundation (SNF).
We are grateful to CERN for their hospitality 
and thank R.~Oliveira and the TS/DEM group, where several of the components
our of detector were manufactured. We also thank
the RD51 Collaboration for useful discussions and suggestions.

\bibliographystyle{elsarticle-num}

\end{document}